\documentclass[letterpaper, 10 pt, conference]{ieeeconf}  

\IEEEoverridecommandlockouts                              
                                                         
\usepackage{xcolor}
\usepackage{cite}
\usepackage{amsmath,amssymb,amsfonts}
\usepackage{algorithmic}
\usepackage{algorithm}
\usepackage{graphicx}
\usepackage{subfig}
\usepackage{textcomp}

\graphicspath{{pics/}}

\usepackage{mathtools}

\usepackage[acronym, style=alttree, toc=true, shortcuts, xindy, nomain, nonumberlist]{glossaries}

\usepackage{cleveref}

\usepackage{hyphenat}
\usepackage{bm}

\usepackage{siunitx}  

\newcommand{\lr}[1]{\left(#1\right)} 

\DeclarePairedDelimiterXPP\onenorm[1]{}\lVert\rVert{_1}{\ifblank{#1}{\:\cdot\:}{#1}} 
\DeclarePairedDelimiterXPP\twonorm[1]{}\lVert\rVert{_2}{\ifblank{#1}{\:\cdot\:}{#1}} 

\newtheorem{theorem}{Theorem}
\newtheorem{assumption}{Assumption}

\newtheorem{definition}{Definition}

\newglossary{symbols}{sym}{sbl}{List of Symbols}
\newglossary{notation}{not}{nt}{Notation}

\glsaddkey{dimension}{\glsentrytext{\glslabel}}{\glsentryd}{\GLsentryd}{\glsd}{\Glsd}{\GLSd}
\glsaddkey{body}{\glsentrytext{\glslabel}}{\glsentrybody}{\GLsentrybody}{\glsbody}{\Glsbody}{\GLSbody}

\newacronym{MPC}{MPC}{Model Predictive Control}
\newacronym{MPCa}{MPC algorithm}{MPC algorithm}
\newacronym{RMPC}{RMPC}{Robust Model Predictive Control}
\newacronym{SMPC}{SMPC}{Stochastic Model Predictive Control}
\newacronym{SCMPC}{SCMPC}{Scenario Model Predictive Control}
\newacronym{MILP}{MILP}{Mixed Integer Linear Program}
\newacronym{PIT}{PIT}{Pointwise\hyp{}In\hyp{}Time}
\newacronym{POMDP}{POMDP}{Partially Observable Markov Decision Process}
\newacronym{MDP}{MDP}{Markov Decision Process}
\newacronym{KKT}{KKT}{Karush\hyp{}Kuhn\hyp{}Tucker}
\newacronym{ev}{EV}{ego vehicle}
\newacronym{tv}{TV}{target vehicle}
\newacronym{cog}{CoG}{center of gravity}
\newacronym{ol}{OL}{Open-Loop}
\newacronym{cl}{CL}{Closed-Loop}
\newacronym{ocp}{OCP}{Optimal Control Problem}
\newacronym{pog}{PG}{\textit{Probabilistic Grid}}
\newacronym{bog}{BG}{\textit{Binary Grid}}
\newacronym{lk}{LK}{lane keeping}
\newacronym{lc}{LC}{lane changing}
\newacronym{og}{OG}{Occupancy Grid}

\title{\LARGE \bf
Model Predictive Control with Models of Different Granularity \\and a Non-uniformly Spaced Prediction Horizon
}

\author{Tim~Br\"udigam,~Daniel~Prader,~Dirk~Wollherr,~and~Marion~Leibold
\thanks{The authors are with the Chair of Automatic Control Engineering at the Technical University of Munich, Arcisstrasse 21, 80333 Munich, Germany.
{\tt\small \{tim.bruedigam; daniel.prader; dw; marion.leibold\}@tum.de}} 
}

\begin{document}

\maketitle
\thispagestyle{empty}
\pagestyle{empty}

\begin{abstract}
Horizon length and model accuracy are defining factors when designing a Model Predictive Controller. While long horizons and detailed models have a positive effect on control performance, computational complexity increases. As predictions become less precise over the horizon length, it is worth investigating a combination of different models and varying time step size. Here, we propose a Model Predictive Control scheme that splits the prediction horizon into two segments. A detailed model is used for the short-term prediction horizon and a simplified model with an increased sampling time is employed for the long-term horizon. This approach combines the advantage of a long prediction horizon with a reduction of computational effort due to a simplified model and less decision variables. The presented Model Predictive Control is recursively feasible. A simulation study demonstrates the effectiveness of the proposed method: employing a long prediction horizon with advantages regarding computational complexity.
\end{abstract}

\section{Introduction}
\label{sec:introduction}

\vspace{-11.6cm}
\mbox{\small 
This~work~has~been~accepted~to~the~2021~American~Control~Conference.}
\vspace{10.8cm}

\vspace{-11.2cm}
\mbox{\small The~published~version~is~available~at~https://doi.org/10.23919/ACC50511.2021.9482617}
\vspace{10.4cm}

\gls{MPC} iteratively solves an optimal control problem on a finite prediction horizon. Over this horizon, a cost function is minimized and constraints are satisfied, given a system prediction model \cite{RawlingsMayneDiehl2017, Mayne2014}. When designing an MPC controller, horizon length and model accuracy need to be chosen to fit the control task. 

While long horizons and detailed models improve the prediction, this also results in increased computational effort. Detailed models provide precise short-term predictions, however, even small model inaccuracies can accumulate over a long prediction horizon, leading to the question how detailed a long-term prediction model needs to be. In certain applications, it is useful to plan precisely for the short-term future while only roughly planning the long-term future. Consider the task of controlling an automated vehicle. Whereas precise planning with a detailed prediction model is fundamental for the immediate future, long-term aims, such as smart lane decisions, do not require a detailed prediction model. However, accounting for long-term aims is still beneficial, for example switching to the right lane early in dense traffic facilitates a right turn later.

Various approaches have been suggested to tackle the issue of long prediction horizons and model accuracy. Hierarchical MPC methods \cite{Scattolini2009}, use multiple MPC levels with varying complexity. However, the optimal control problems are solved individually, e.g., a high level regulator with slow time scale on a reduced order model and a low level regulator with fast time scale in \cite{FarinaZhangScattolini2018}. Hierarchical MPC schemes are especially popular for chemical applications where different time scales are present.

MPC with move blocking \cite{CagienardEtalMorari2004, GondhalekarImura2010, ShekharManzie2015} provides an approach to reduce the number of decision variables within the optimal control problem. Regarding input move blocking, certain inputs along the prediction horizon are set equal to previous input values. However, shifting the blocked inputs when solving the optimal control problem is an issue. A flexible move blocking strategy was proposed in \cite{SchwickartEtalBezzaoucha2016}, adapting the blocking when relevant.

In \cite{BaethgeLuciaFindeisen2016} an MPC scheme is proposed, which uses two different models over the prediction horizon. A detailed model for the short-term horizon is combined with an approximated, coarse prediction model for the long-term horizon. A robust MPC approach \cite{MayneSeronRakovic2005} is chosen for the long-term horizon to account for model mismatch. While recursive feasibility is guaranteed, stability is not shown. In~\cite{BruedigamEtalLeibold2020b} the approach in~\cite{BaethgeLuciaFindeisen2016} is extended. A robust MPC approach with a detailed model is combined with a stochastic MPC method \cite{Mesbah2016} and a simplified model. The approach in \cite{BruedigamEtalLeibold2020b} allows safe planning for the short-term future while still accounting for increased uncertainty in the long-term future with probabilistic constraints, i.e., chance constraints. In \cite{ZanelliEtalDiehl2017} a real-time iteration scheme for nonlinear MPC is presented, where constraints in the later part of the prediction horizon are replaced by logarithmic barriers. 

A different approach to reduce computational effort is presented in \cite{TippettTanBao2014, TanTippettBao2015}. Only a single prediction model is employed over the prediction horizon, however, the sampling time is varied, resulting in an MPC scheme with a non-uniformly spaced optimization horizon. The sampling time increases along the prediction horizon, allowing to extend the time covered by the horizon while keeping the amount of decision variables constant. While stability, based on dissipativity, is shown, recursive feasibility is not addressed. Both MPC with models of different granularity and MPC with a non-uniformly spaced optimization horizon exploit less detailed planning for the long-term future in order to reduce computational complexity. However, both methods only focus on one specific aspect of reducing the computational complexity.

In this paper, we propose an MPC scheme that combines the approaches of \cite{BaethgeLuciaFindeisen2016} and \cite{TanTippettBao2015}. The prediction horizon is divided into two segments. A detailed model with relatively small sampling time is combined with an approximated, coarse model and larger sampling time. This allows to use benefits of both individual methods. Computational effort due to model complexity is reduced by utilizing the simplified model for the long-term horizon. Furthermore, the time covered by the prediction horizon is extended by choosing larger sampling times while the amount of decision variables remains constant. Recursive feasibility of the proposed method is guaranteed.

The presented approach is beneficial for tasks requiring precise control for the short-term future, ensured by a small sampling time and a detailed model, where additionally long-term, coarse planning is advantageous. The long-term planning allows to incorporate long-term goals into short-term planning in such a way that it does not compromise the required short-term precision, i.e., the cost-to-go is improved. We show the effectiveness of the proposed method in a brief vehicle avoidance simulation. 

The paper is structured as follows. In Section~\ref{sec:problem} the considered systems are introduced and MPC with models of different granularity as well as MPC with a non-uniformly spaced horizon are summarized. The proposed method, MPC with models of different granularity in combination with a non-uniformly spaced horizon, is presented in Section~\ref{sec:method}. A discussion is given in Section~\ref{sec:discussion}. A simulation study is shown in Section~\ref{sec:results}, followed by conclusive remarks in Section~\ref{sec:conclusion}.

\section{Problem Setup}
\label{sec:problem}

Similar to \cite{BaethgeLuciaFindeisen2016} we consider two nonlinear, discrete-time system models

\vspace{-5mm}
\begin{minipage}[t]{0.45\columnwidth}
\begin{subequations}\label{eq:model_bei_det_2}
\begin{align}
\bm{x}^+ = \bm{f}(\bm{x},\bm{u})\\
\text{s.t.}\quad \bm{x} \in \mathbb{X}\\
\bm{u} \in \mathbb{U}
\end{align}
\end{subequations}
\end{minipage}
\begin{minipage}[t]{0.45\columnwidth}
\begin{subequations}\label{eq:model_bei_cor_2}
\begin{align}
\bm{z}^+ = \bm{g}(\bm{z},\bm{v})\\
\text{s.t.}\quad \bm{z} \in \mathbb{Z}\\
\bm{v} \in \mathbb{V}
\end{align}
\end{subequations}
\end{minipage}\\

\noindent
with inputs $ \bm{u}\in\mathbb{R}^{n_u} $ and $ \bm{v}\in\mathbb{R}^{n_v}$, and states $ \bm{x}\in\mathbb{R}^{n_x} $ and $ \bm{z}\in\mathbb{R}^{n_z} $, where $\bm{x}^+$ and $\bm{z}^+$ denote the states at the next time step. Here, model \eqref{eq:model_bei_cor_2} is considered to be an approximation of model \eqref{eq:model_bei_det_2}. The state and input constraints are given by the state and input constraint sets $\mathbb{X}$, $\mathbb{U}$, and $\mathbb{Z}$, $\mathbb{V}$, respectively.

We now extend the models \eqref{eq:model_bei_det_2} and \eqref{eq:model_bei_cor_2} including different sampling times

\vspace{-5mm}
\begin{minipage}[t]{0.45\columnwidth}
\begin{subequations}\label{eq:model_bei_det_2st}
\begin{align}
\bm{x}^+ = \bm{f}(\bm{x},\bm{u},\Delta t_\text{s})\\
\text{s.t.}\quad \bm{x} \in \mathbb{X}\\
\bm{u} \in \mathbb{U}
\end{align}
\end{subequations}
\end{minipage}
\begin{minipage}[t]{0.45\columnwidth}
\begin{subequations}\label{eq:model_bei_cor_2st}
\begin{align}
\bm{z}^+ = \bm{g}(\bm{z},\bm{v},\Delta t_\text{f})\\
\text{s.t.}\quad \bm{z} \in \mathbb{Z}\\
\bm{v} \in \mathbb{V}
\end{align}
\end{subequations}
\end{minipage}\\

\noindent
where $\Delta t_\text{s}$ and $\Delta t_\text{f}$ are the sampling times for the respective prediction models. The two models are linked given a projection function as defined in \cite{BaethgeLuciaFindeisen2016}.

\begin{assumption}
\label{ass:projection}
There exists a surjective projection function $ \text{Proj} : \mathbb{R}^{n_x} \times \mathbb{R}^{n_u} \rightarrow \mathbb{R}^{n_z} \times \mathbb{R}^{n_v} $ mapping the state and input $\bm{x}$ and $\bm{u}$ of the detailed model to the state $\bm{z}$ and $\bm{v}$ of the coarse model \cite{BaethgeLuciaFindeisen2016}. 
\end{assumption}

Ideally, the constraint sets $\mathbb{Z}$ and $\mathbb{V}$ of the coarse model are computed using this projection function, i.e., $ (\mathbb{Z},\mathbb{V})  = \text{Proj} (\mathbb{X},\mathbb{U}) $. If the main focus is to improve the cost-to-go and feasibility issues occur, the constraint sets $\mathbb{Z}$ and $\mathbb{V}$ may be chosen to depend more loosely on $\mathbb{X}$ and $\mathbb{U}$.

In the following, two concepts are presented, MPC with models of different granularity \cite{BaethgeLuciaFindeisen2016} and MPC with a non-uniformly spaced prediction horizon \cite{TanTippettBao2015}, which will later be combined.

\subsection{MPC with Models of Different Granularity}

We will briefly summarize MPC with models of different granularity, based on the idea presented in \cite{BaethgeLuciaFindeisen2016}, which relies on splitting the original MPC prediction horizon $N$ into two parts. A short-term horizon ranges from $k=0$ to $k = k_\text{s}$ and a long-term horizon starts at $k = k_\text{s}$ and ends at $k = k_\text{f} = N$. For the short-term prediction a detailed model \eqref{eq:model_bei_det_2} is used, whereas the long-term prediction is based on a coarse model~\eqref{eq:model_bei_cor_2}. 

This results in an MPC cost function
\begin{IEEEeqnarray}{rl}
\min_{\substack{\{\bm{u}_0,...,\bm{u}_{k_{\text{s}}}\},\\ \{\bm{v}_{k_{\text{s}}},...,\bm{v}_{k_{\text{f}}-1}\} }} 
\Biggl(
&\sum_{k=0}^{k_{\text{s}}-1} l_{\text{s}}(\bm{x}_k,\bm{u}_k)+V_{\text{s}}(\bm{x}_{k_{\text{s}}}) \nonumber \\
&+\sum_{k=k_{\text{s}}}^{k_{\text{f}}-1} l_{\text{f}}(\bm{z}_k,\bm{v}_k)+V_{\text{f}}(\bm{z}_{k_{\text{f}}})
\Biggl)
\end{IEEEeqnarray}
with stage cost functions $l_{\text{s}}(\bm{x}_k,\bm{u}_k)$, $l_{\text{f}}(\bm{z}_k,\bm{v}_k)$ and terminal cost functions $V_{\text{s}}(\bm{x}_{k_{\text{s}}})$, $V_{\text{f}}(\bm{x}_{k_{\text{f}}})$. Constraints of the detailed model, according to \eqref{eq:model_bei_det_2}, are considered for $k = 0, ..., k_\text{s}-1$ and constraints of the coarse model \eqref{eq:model_bei_cor_2} are required to hold for $k = k_\text{s}, ..., k_\text{f}-1$, with the terminal constraint $\bm{z}(k_\text{f}) \in \mathbb{Z}_\text{f} \subseteq \mathbb{Z}$. A projection 
\begin{IEEEeqnarray}{c}
(\bm{z}_{k_{\text{s}}},\bm{v}_{k_{\text{s}}})=\text{Proj}(\bm{x}_{k_{\text{s}}},\bm{u}_{k_{\text{s}}})\subseteq \mathbb{Z} \times \mathbb{V}
\end{IEEEeqnarray}
links the two models at $k = k_\text{s}$.

\subsection{MPC with Non-uniformly Spaced Horizon}
\label{sec:NUSH}

In \cite{TippettTanBao2014, TanTippettBao2015} an approach is presented to use varying sampling time within the MPC prediction horizon, allowing to extend the horizon time without increasing the number of decision variables. While multiple different time steps $\Delta t_j$ can be used, using only two different time steps is often satisfactory, $\Delta t_1$ for the short-term horizon and $\Delta t_2$ for the long-term horizon. Typically, different $\Delta t_j$ are chosen such that $\Delta t_{j+1} > \Delta t_j$, i.e., sampling time increases over the prediction horizon.

While constraints can be chosen independently for each horizon segment, the different sampling time needs to be accounted for in the quadratic cost function with weighting matrices $\bm{Q}$ and $\bm{R}$ for states and inputs, respectively. This is fundamental in order to penalize horizon segments with larger $\Delta t_j$ values equally, even though the relative number of predicted states and inputs affecting the cost is less compared to horizon segments with small $\Delta t_j$. This is achieved by adapting the original weighting matrices $\bm{Q}$ and $\bm{R}$ to individual weighting matrices $\bm{Q}_j$ and $\bm{R}_j$ for each horizon segment according to
\begin{IEEEeqnarray}{c}
\bm{Q}_j=\bm{Q}\frac{\Delta t_j}{\Delta t_1}, ~~ \bm{R}_j=\bm{R}\frac{\Delta t_j}{\Delta t_1}. \label{eq:qadapt}
\end{IEEEeqnarray}

The presented approaches, MPC with models of different granularity and MPC with a non-uniformly spaced horizon, are now combined.

\section{Method}
\label{sec:method}

In this section, the MPC optimal control problem will be presented, which includes models of different granularity and a non-uniformly spaced prediction horizon. Recursive feasibility of the approach is shown, followed by a discussion.

\subsection{Optimal Control Problem}

\begin{figure}
    \centering
    \includegraphics[width = \columnwidth]{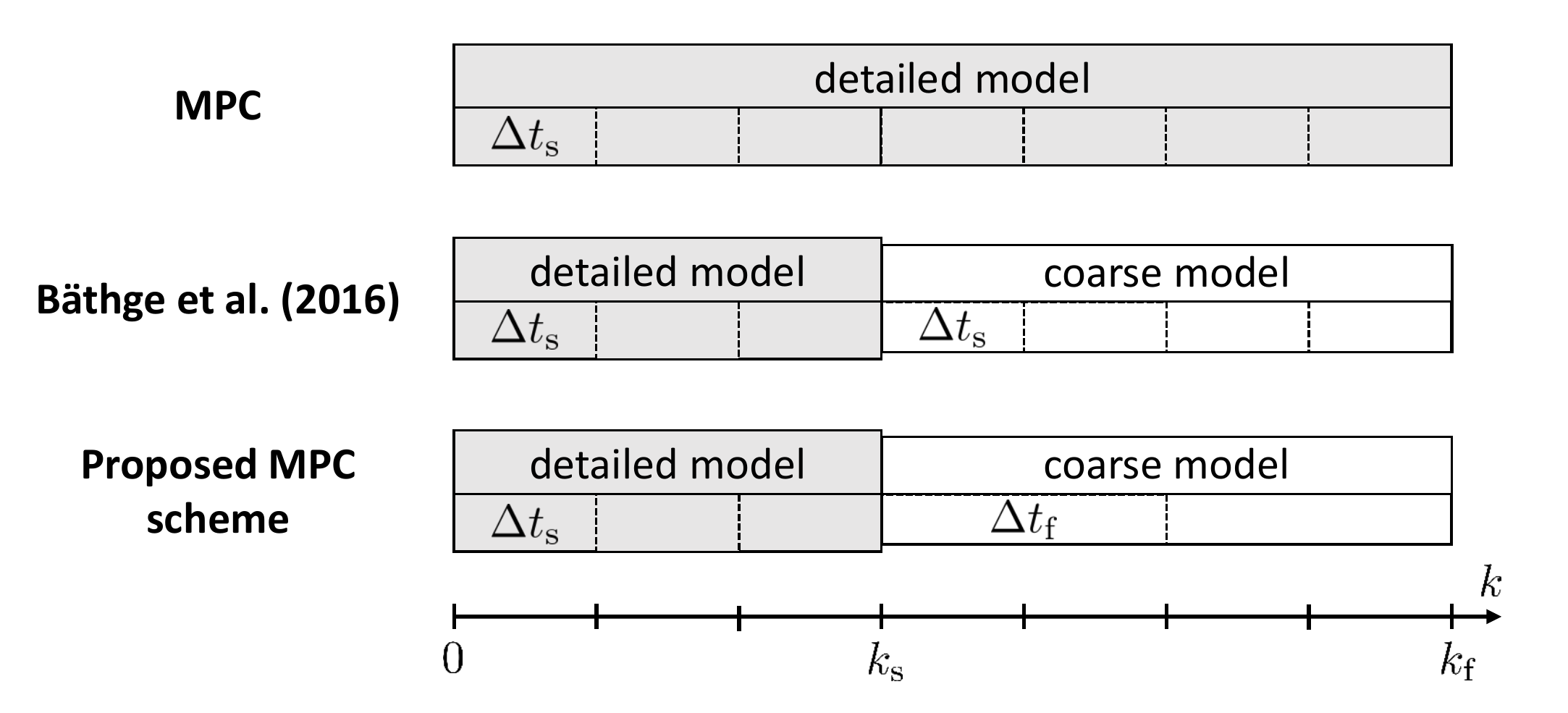}
    \caption{MPC scheme comparison}
    \label{fig:method}
\end{figure}

Given the methods presented in Section \ref{sec:problem}, we combine a detailed model and small sampling time for the short-term horizon with a coarse model and larger sampling time for the long-term horizon. The idea is displayed in Figure \ref{fig:method}. 
The MPC optimal control problem is given by
\begin{IEEEeqnarray}{l}
\label{eq:mixedMPC_OCP} \IEEEyesnumber 
\min_{\substack{\{\bm{u}_0,...,\bm{u}_{k_{\text{s}}}\},\\ \{\bm{v}_{k_{\text{s}}},...,\bm{v}_{k_{\text{f}}-1}\} }} 
\Biggl(
\sum_{k=0}^{k_{\text{s}}-1} l_{\text{s}}(\bm{x}_k,\bm{u}_k,\Delta t_\text{s})+V_{\text{s}}(\bm{x}_{k_{\text{s}}}) \IEEEeqnarraynumspace \nonumber \\
\hspace{20mm}+\sum_{k=k_{\text{s}}}^{k_{\text{f}}-1} l_{\text{f}}(\bm{z}_k,\bm{v}_k,\Delta t_\text{f})+V_{\text{f}}(\bm{z}_{k_{\text{f}}})
\Biggl) \IEEEeqnarraynumspace \IEEEyessubnumber \label{eq:MPC_cost}\\
\text{s.t. }~ 
\hspace{0.2mm} \bm{x}_{k+1}=\bm{f}(\bm{x}_k,\bm{u}_k,\Delta t_\text{s}), \IEEEyessubnumber \label{eq:MPC_x0} \\
\hspace{6.6mm} \bm{x}_k\in \mathbb{X},\hspace{10mm} \forall k = 1,...,k_\text{s}-1,  \IEEEyessubnumber \\
\hspace{6.6mm} \bm{u}_k\in \mathbb{U},\hspace{10mm} \forall k = 0,...,k_\text{s}, \IEEEyessubnumber \\
\hspace{6.6mm} \bm{x}_{k_\text{s}}\in \mathbb{X}_{\text{CI}} \subseteq \mathbb{X}, \IEEEyessubnumber \label{eq:MPC_CIS} \\
\hspace{6.6mm} (\bm{z}_{k_\text{s}},\bm{v}_{k_\text{s}})=\text{Proj}\lr{\bm{x}_{k_\text{s}},\bm{u}_{k_\text{s}}}\subseteq \mathbb{Z} \times \mathbb{V},  \IEEEyessubnumber \label{eq:MPC_Proj}\\
\hspace{6.6mm} \bm{z}_{k+1}=\bm{g}(\bm{z}_k,\bm{v}_k,\Delta t_\text{f}), \IEEEyessubnumber  \\
\hspace{6.6mm} \bm{z}_k\in \mathbb{Z},\hspace{10mm} \forall k = k_\text{s}+1,...,k_\text{f},  \IEEEyessubnumber \\
\hspace{6.6mm} \bm{v}_k\in \mathbb{V},\hspace{10mm} \forall k = k_\text{s},...,k_\text{f}-1 \IEEEyessubnumber \label{eq:MPC_vV}
\end{IEEEeqnarray}
with the current system state $\bm{x}_0$, the standard sampling time $\Delta t_\text{s}$ and a larger sampling time $\Delta t_\text{f} > \Delta t_\text{s}$, as well as a control invariant set $\mathbb{X}_{\text{CI}}$. The input $\bm{u}_{k_\text{s}}$ is necessary to evaluate \eqref{eq:MPC_Proj}.

The stage costs $l_{\text{s}}(\bm{x}_k,\bm{u}_k,\Delta t_\text{s})$ and $l_{\text{f}}(\bm{z}_k,\bm{v}_k,\Delta t_\text{f})$ depend on the respective sampling time. For a quadratic cost function a similar approach to~\cite{TanTippettBao2015}, presented in Section~\ref{sec:NUSH}, can be applied to adapt the weighting matrices, based on $\Delta t_\text{s}$ and $\Delta t_\text{f}$. The terminal cost functions are given by $V_{\text{s}}$ and $V_{\text{f}}$.

The proposed MPC scheme allows to apply an accurate prediction model with small sampling time for precise short-term predictions, while still considering long-term aims, with a less accurate long-term prediction. In the following, recursive feasibility of the MPC scheme is shown.

\subsection{Recursive Feasibility}

One of the fundamental challenges for MPC is to be able to guarantee recursive feasibility, as the optimal control problem needs to be solved iteratively. Here, it is not possible to apply standard MPC theory, e.g., shifting the previous input sequence and a control invariant terminal constraint, as the sampling time changes for the long-term horizon, i.e., $\Delta t_\text{s} < \Delta t_\text{f}$. In the following, an input $\bm{u}_{k|t}$ indicates the input for the prediction step $k$ at time step $t$. This similarly holds for $\bm{x}$, $\bm{z}$, and $\bm{v}$.

\begin{definition}
MPC is recursively feasible if a feasible input 
\begin{IEEEeqnarray}{c}
\label{eq:Ut}
[\bm{U}_t, \bm{V}_t] = [\bm{u}_{0|t},, ..., \bm{u}_{k_\text{s}|t},\bm{v}_{k_\text{s}|t}, ..., \bm{v}_{(k_\text{f}-1)|t}]
\end{IEEEeqnarray}
at time step $t$, satisfying \eqref{eq:MPC_x0}-\eqref{eq:MPC_vV}, guarantees that the MPC optimal control problem is feasible at time step $t+1$, i.e., a solution $[\bm{U}_{t+1}, \bm{V}_{t+1}]$ exists.
\end{definition}

To prove recursive feasibility, the initial optimal control problem must be feasible.

\begin{assumption}
\label{ass:init}
The optimal control problem \eqref{eq:mixedMPC_OCP} is initially feasible, i.e., a solution $[\bm{U}_t, \bm{}V_t]$, according to \eqref{eq:Ut}, exists for $t=0$.
\end{assumption}

Additionally, the difference in sampling time must be considered.
\begin{assumption}
\label{ass:CIS}
A control invariant set $\mathbb{Z}_{\text{CI}}$ can be obtained, so that for all $\bm{x}_k \in \mathbb{X}_{\text{CI}}$, it follows that $\bm{z}_k \in \mathbb{Z}_{\text{CI}}$.
\end{assumption}

This assumption implicates the following. If a control invariant set for a state $x_k$ with model \eqref{eq:model_bei_det_2st} exists, a control invariant set also exists for the corresponding state $\bm{z}_k$ with model \eqref{eq:model_bei_cor_2st}, given the different sampling time. 

\begin{theorem}\label{th:recfeas}
The MPC optimal control problem \eqref{eq:mixedMPC_OCP} is recursively feasible if Assumptions \ref{ass:projection}, \ref{ass:init}, and \ref{ass:CIS} hold.
\end{theorem}

The theorem is proved by showing that a feasible $[\bm{U}_{t+1}, \bm{V}_{t+1}]$ exists, given a feasible $[\bm{U}_{t}, \bm{V}_{t}]$.

\begin{proof}
Due to Assumption \ref{ass:init}, an input $[\bm{U}_t, \bm{V}_t] = [\bm{u}_{0|t}, ..., \bm{u}_{k_\text{s}|t}, \bm{v}_{k_\text{s}|t}, ..., \bm{v}_{(k_{\text{f}}-1)|t}]$ exists. 
First, the focus is on the short-term horizon, using $\bm{U}_t$. Shifting the initial segment of inputs $[\bm{u}_{0|t}, ..., \bm{u}_{(k_{\text{s}}-1)|t}]$ by one step yields the series of inputs $[\bm{u}_{0|(t+1)}, ..., \bm{u}_{(k_{\text{s}}-2)|(t+1)}]$, as the previous input sequence remains feasible for step $t+1$. According to \eqref{eq:MPC_CIS}, $\bm{x}_{k_\text{s}|t}$ lies in the control invariant set $\mathbb{X}_\text{CI}$, therefore, $\bm{x}_{(k_{\text{s}}-1)|(t+1)} \in \mathbb{X}_\text{CI}$ and an input $\bm{u}_{(k_{\text{s}}-1)|(t+1)}$ exists such that $\bm{x}_{k_\text{s}|(t+1)} \in \mathbb{X}_\text{CI}$. This implies an input $\bm{u}_{k_\text{s}|(t+1)}$ exists, yielding the input sequence $\bm{U}_{t+1} = [\bm{u}_{0|(t+1)}, ..., \bm{u}_{(k_\text{s}-1)|(t+1)}, \bm{u}_{k_\text{s}|(t+1)}]$. 

Next, the long-term horizon is considered. Given Assumption \ref{ass:CIS}, $\bm{z}_{k_\text{s}} \in \mathbb{Z}_\text{CI}$ and $\bm{v}_{k_\text{s}}$ exists. As $\mathbb{Z}_\text{CI}$ is a control invariant set, an input sequence $\bm{V}_{t+1} = [\bm{v}_{k_\text{s}|(t+1)}, ..., \bm{v}_{(k_{\text{f}}-1)|t}]$ exists, yielding $[\bm{U}_{t+1}, \bm{V}_{t+1}] = [\bm{u}_{0|(t+1)}, ..., \bm{u}_{k_\text{s}|(t+1)}, \bm{v}_{k_\text{s}|(t+1)}, ..., v_{(k_{\text{f}}-1)|(t+1)}]$.

Therefore, the MPC optimal control problem \eqref{eq:mixedMPC_OCP} is recursively feasible.
\end{proof}

Note that $\bm{u}_{k_{\text{s}}|t}$ is not part of the cost function \eqref{eq:MPC_cost} and does not guarantee $\bm{x}_{(k_{\text{s}}+1)|t} \in \mathbb{X}_\text{CI}$, but is necessary to evaluate~\eqref{eq:MPC_Proj}.

\section{Discussion}
\label{sec:discussion}

The presented method divides the prediction horizon into two segments. Multiple segments extending the original horizon with different simpler models and larger sampling times are also possible. However, the effort of designing and setting up multiple segments could be higher than the resulting benefit.

The proposed approach can be interpreted and applied in two ways with respect to standard MPC: extending or splitting the horizon. In a first interpretation, the second horizon segment is regarded as an extended horizon compared to the standard MPC horizon. This allows longer predictions, while the computational effort is only slightly increased due to a simplified model and larger sampling times. A second interpretation is as follows. The time span covered by the prediction horizon is equal for standard MPC and the proposed method. But computational complexity is reduced as the detailed model is only employed for the short-term prediction and less decision variables are used, given the non-uniformly spaced horizon.

In contrast to \cite{BaethgeLuciaFindeisen2016} and other literature, in the optimal control problem \eqref{eq:mixedMPC_OCP} the control invariant set $\mathbb{X}_\text{CI}$ is at the end of the first horizon segment ($\bm{x}_{k_\text{s}}$). This is necessary to guarantee recursive feasibility. If the control invariant set were at the end of the overall horizon, recursive feasibility could not be guaranteed, as the different sampling times in the short- and long-term horizon do not allow standard MPC theory to guarantee recursive feasibility, i.e., reusing the shifted horizon for the next time step is not possible. This is similar to guaranteeing stability in MPC with a non-uniformly spaced horizon \cite{TanTippettBao2015}. However, if we interpret the proposed method as an approach, which extends the standard horizon with a long-term horizon to improve the prediction at only slightly increased computational effort, it is suitable to place a control invariant set at the end of the short-term horizon. 

The predicted states in the long-term horizon do not affect recursive feasibility, as $\bm{x}_{k_\text{s}} \in \mathbb{X}_\text{CI}$ ensures that the optimal control problem remains recursively feasible. The long-term horizon is considered as an improvement for the cost-to-go. Therefore, the constraints for the long-term horizon do not necessarily have to exactly match the constraints of the short-term horizon. A set $\mathbb{Z}_\text{CI}$ must still be provided, however, to ensure that the proof of Theorem~\ref{th:recfeas} remains valid. However, there is a certain degree of freedom to select $\mathbb{Z}_\text{CI}$.

As stability was not yet shown for MPC with models of different granularity, the focus of this work was to first guarantee recursive feasibility, which is guaranteed for MPC with models of different granularity but not for MPC with a non-uniformly spaced horizon. Dissipativity theory could be of interest, similar to the stability guarantee in \cite{TanTippettBao2015}, when investigating stability for the proposed method.

In \cite{BaethgeLuciaFindeisen2016} a robust MPC scheme was employed for the long-term horizon to address consistency of the models. While this was omitted here to focus on the combination of different models and varying sampling time, a robust MPC scheme could be applied for the long-term prediction together with additive noise to the coarse system model \eqref{eq:model_bei_cor_2st}.

It is also important to note that not any simplified model is suitable to be combined with a detailed model. It must be possible to find a projection function, which is more likely if the coarse model is a reduced model of the detailed model. Finding a reduced model for a detailed nonlinear model is challenging. However, for detailed linear models, it is often straightforward to obtain a reduced model that ensures that Assumption~\ref{ass:projection} is fulfilled. An example will be addressed in the following simulation study.

\section{Simulation Study}
\label{sec:results}

\begin{figure}
    \centering
    \includegraphics[width = 0.97\columnwidth]{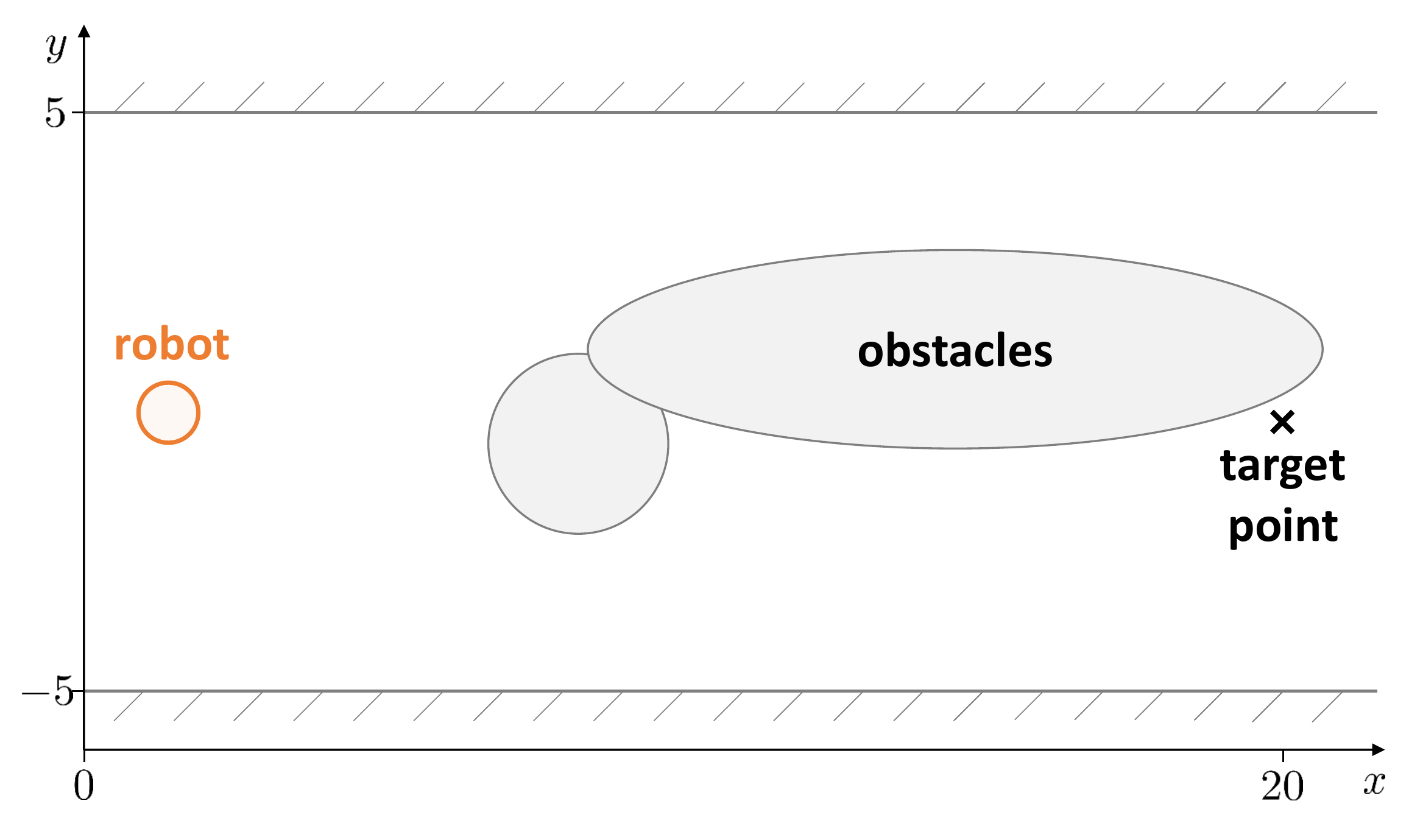}
    \caption{Simulation scenario}
    \label{fig:scenario}
\end{figure}

We evaluate the proposed MPC method in a setting similar to the one described in \cite{BaethgeLuciaFindeisen2016}. A mobile robot is steered along a path with obstacles, as illustrated in Figure~\ref{fig:scenario}. The aim is to reach the target point while avoiding obstacles. 
All quantities are given in SI units. The simulations were carried out in MATLAB with the $\mathrm{fmincon}$ solver on a standard desktop computer.

\subsection{System Models}

Two system models are considered, where the nonlinearity is found in the constraints. The detailed model is given by
\begin{IEEEeqnarray}{c}
\bm{x}^+=
\begin{bmatrix}
1 & \Delta t & 0 & 0 \\ 0 & 1 & 0&0\\0&0&1&\Delta t\\0&0&0&1
\end{bmatrix}
\bm{x}+
\begin{bmatrix}
0 & 0 \\ \Delta t \frac{1}{m}&0\\0&0\\0&\Delta t \frac{1}{m} 
\end{bmatrix}
\bm{u}
\IEEEeqnarraynumspace \label{eq:sim_model_det} \end{IEEEeqnarray}
with state vector $\bm{x} = [p_x, v_x, p_y, v_y]^\top$, input $\bm{u} = [F_x, F_y]^\top$, sampling time $\Delta t$, and mass $m = 0.5$. The state vector consists of $x-$ and $y-$position $p_x$ and $p_y$, as well as $x-$ and $y-$velocity $v_x$ and $v_y$, the inputs are forces in $x-$ and $y-$direction, $F_x$ and $F_y$. The following constraints, $\mathbb{X}$ and $\mathbb{U}$, are employed for states and inputs
\begin{IEEEeqnarray}{rcl}
\label{eq:constraints_det} \IEEEyesnumber 
-3 \leq &v_x &\leq 3 \IEEEyessubnumber \\
-5 \leq &p_y &\leq 5 \IEEEyessubnumber \\
-3 \leq &v_y &\leq 3 \IEEEyessubnumber \\
-3 \leq &F_x &\leq 3 \IEEEyessubnumber \\
-0.5 \leq &F_x &\leq 0.5. \IEEEyessubnumber
\end{IEEEeqnarray}
The control invariant set $\mathbb{X}_{\text{CI}}$ is given by 
\begin{IEEEeqnarray}{c}
v_x = 0,~~ v_y = 0,~~ -5 \leq p_y \leq 5. \label{eq:CIconstraint}
\end{IEEEeqnarray}
This ensures that at the end of the first horizon segment, the robot can come to a standstill, which avoids any constraint violations.

The approximated, coarse model, based on \eqref{eq:sim_model_det}, is given by
\begin{IEEEeqnarray}{c}
\bm{z}^+=
\begin{bmatrix}
1 & 0 \\0&1
\end{bmatrix}
\bm{z}+
\begin{bmatrix}
\Delta t&0\\0&\Delta t 
\end{bmatrix}
\bm{v} \label{eq:sim_model_cor}
\end{IEEEeqnarray}
with state $z = [p_x, p_y]^\top$ and input $u = [v_x, v_y]^\top$. The models \eqref{eq:sim_model_det} and \eqref{eq:sim_model_cor} are linked by the projection matrix
\begin{IEEEeqnarray}{c}
\begin{pmatrix}
\bm{z}\\\bm{v}
\end{pmatrix}=
\text{Proj}\left(
\begin{pmatrix}
\bm{x}\\\bm{u}
\end{pmatrix}\right)=
\begin{bmatrix}
1&0&0&0&0&0\\0&0&1&0&0&0\\0&1&0&0&0&0\\0&0&0&1&0&0
\end{bmatrix}
\begin{pmatrix}
\bm{x}\\\bm{u}
\end{pmatrix}. \IEEEeqnarraynumspace \label{eq:sim_projection}
\end{IEEEeqnarray}
The coarse model is subject to constraints $\mathbb{Z}$ and $\mathbb{V}$ similar to \eqref{eq:constraints_det}, i.e.,
\begin{IEEEeqnarray}{rcl}
\label{eq:constraints_cor} \IEEEyesnumber 
-5 \leq &p_y &\leq 5 \IEEEyessubnumber \\
-3 \leq &v_x &\leq 3 \IEEEyessubnumber \\
-3 \leq &v_y &\leq 3, \IEEEyessubnumber 
\end{IEEEeqnarray}
where the control invariant set $\mathbb{Z}_{\text{CI}}$ is defined as in~\eqref{eq:CIconstraint}.

Additionally, obstacles according to Figure~\ref{fig:scenario} are considered as ellipsoidal constraints in both the detailed and the coarse model. Given the ellipse equation 
\begin{IEEEeqnarray}{C}
\lr{\frac{p_x-x^*}{a}}^2 + \lr{\frac{p_y-y^*}{b}}^2 \leq 1
\end{IEEEeqnarray}
with ellipse parameters $a$ and $b$ and origin offset $[x^*, y^*]$, we consider the two overlapping obstacles with parameters $[a_1, b_1, x^*_1, y_1^*] = [1.5, 1.5, 10, -0.1]$ and $[a_2, b_2, x^*_2, y_2^*] = [5, 1.4, 15.2, 1.3]$. 

The position of the two obstacles allows to analyze the benefit of a longer prediction horizon. Passing the obstacles above results in a longer path. However, the circular obstacle is positioned in such a way ($y_1^* < 0$) that it is more rewarding to pass it above. A longer prediction horizon now allows to choose the path with higher short-term cost, as it has lower cost in the long-term.

\subsection{MPC Schemes}

We compare three MPC setups to evaluate the proposed method: standard MPC, MPC with models of different granularity, and the proposed approach. The standard MPC has a shorter horizon to then show the advantage of using a longer prediction horizon. The aim is to reach the reference point $(p_x, p_y) = (20, 0)$, resulting in the reference states $\bm{x}_\text{ref} = [20, 0, 0, 0]^\top$ and $\bm{z}_\text{ref} = [20, 0]^\top$. The initial state is $\bm{x}_0 = [0,0,0,0]$. All stage costs have the quadratic form
\begin{IEEEeqnarray}{rl}
\IEEEyesnumber
l_\text{s}(\bm{x}_k,\bm{u}_k) &= (\bm{x}_k-\bm{x}_\text{ref})^\top \bm{Q}_\text{s} (\bm{x}_k-\bm{x}_\text{ref}) + \bm{u}_k^\top \bm{R}_\text{s} \bm{u}_k \IEEEeqnarraynumspace \IEEEyessubnumber \label{eq:cost_s} \\
l_\text{f}(\bm{z}_k,\bm{v}_k) &= (\bm{z}_k-\bm{z}_\text{ref})^\top \bm{Q}_\text{f} (\bm{z}_k-\bm{z}_\text{ref}) + \bm{v}_k^\top \bm{R}_\text{f} \bm{v}_k \IEEEeqnarraynumspace \IEEEyessubnumber 
\end{IEEEeqnarray}
with $\bm{Q}_\text{s} = \text{diag}(1, 0, 5, 0)$, $\bm{R}_{\text{s}} = \text{diag}(0.1,0.1)$ and $\bm{Q}_\text{f} = \text{diag}(1, 5)$, $\bm{R}_{\text{f}} = \text{diag}(0.01,0.01)$. While velocities are not penalized in $l_\text{s}(\bm{x}_k,\bm{u}_k)$, they are penalized slightly in $l_\text{f}(\bm{z}_k,\bm{v}_k)$ in order to have a non-zero matrix $\bm{R}_\text{f}$. Terminal cost functions are chosen as $V_{\text{s}}(\bm{x}_k) = (\bm{x}_k-\bm{x}_\text{ref})^\top \bm{Q}_\text{s} (\bm{x}_k-\bm{x}_\text{ref})$ and $V_{\text{f}}(\bm{z}_k) = (\bm{z}_k-\bm{z}_\text{ref})^\top \bm{Q_}\text{f} (\bm{z}_k-\bm{z}_\text{ref})$.

The three controllers have the following characteristics: \\
\textbf{Standard MPC} uses a prediction horizon $N = 10$ with sampling time $\Delta t = 0.2$ for model \eqref{eq:sim_model_det}, constraints \eqref{eq:constraints_det}, and terminal constraints \eqref{eq:CIconstraint}, as well as stage cost $l_\text{s}(\bm{x}_k,\bm{u}_k)$ and $V_{\text{s}}(\bm{x}_{N})$.\\
\textbf{MPC with models of different granularity} uses the horizons $k_\text{s} = 10$ and $k_\text{f} = 16$ with sampling time $\Delta t = 0.2$ with model \eqref{eq:sim_model_det} and constraints \eqref{eq:constraints_det} for the short-term horizon $k_\text{s} = 10$, and model \eqref{eq:sim_model_cor}, constraints \eqref{eq:constraints_cor}, and terminal constraints \eqref{eq:CIconstraint}, between $k_\text{s} = 10$ and the long-term horizon $k_\text{f} = 16$. The stage costs are $l_\text{s}(\bm{x}_k,\bm{u}_k)$ and $l_\text{f}(\bm{z}_k,\bm{v}_k)$ with terminal costs $V_{\text{s}}(\bm{x}_{k_\text{s}})$ and $V_{\text{f}}(\bm{z}_{k_\text{f}})$.\\
\textbf{The proposed MPC scheme} also uses two horizons. For the short-term horizon $k_\text{s} = 10$ with sampling time $\Delta t_1 = 0.2$, model \eqref{eq:sim_model_det} and constraints \eqref{eq:constraints_det} are employed, as well as \eqref{eq:CIconstraint} for the control invariant set $\mathbb{X}_\text{CI}$. Between $k_\text{s} = 10$ and the long-term horizon $k_\text{f} = 16$ the increased sampling time $\Delta t_2 = 0.4$ is chosen with the model \eqref{eq:sim_model_cor}, constraints \eqref{eq:constraints_cor}, and control invariant set $\mathbb{Z}_{\text{CI}}$ according to \eqref{eq:CIconstraint}. Terminal costs $V_{\text{s}}(\bm{x}_{k_\text{s}})$ and $V_{\text{f}}(\bm{z}_{k_\text{f}})$ are used with the stage costs $l_\text{s}(\bm{x}_k,\bm{u_k})$ and 
\begin{IEEEeqnarray}{rl}
l_\text{f}(\bm{z}_k,\bm{v}_k) &= (\bm{z_k}-\bm{z}_\text{ref})^\top \tilde{\bm{Q}}_\text{f} (\bm{z}_k-\bm{z}_\text{ref}) + \bm{v}_k^\top \tilde{\bm{R}}_\text{f} \bm{v}_k \IEEEeqnarraynumspace 
\end{IEEEeqnarray}
with $\tilde{\bm{Q}}_\text{f} = \text{diag}(2, 10)$, $\tilde{\bm{R}}_{\text{f}} = \text{diag}(0.02,0.02)$ according to \eqref{eq:qadapt}. The weights are increased, as the sampling time is larger compared to the short-term horizon, resulting in less states and inputs considered in the cost function.

The main properties of the analyzed MPC schemes (bold font) used for the simulation are summarized in Table~\ref{tab:params}. Properties and results are also provided for further MPC schemes that are not discussed in detail.
\begin{table}[h]
    \centering
    \begin{tabular}{l c c c c c c}
            method & $k_\text{s}$ & $\Delta t_1$ & $k_\text{f} - k_\text{s}$ & $\Delta t_2$ & cost  \\
            \hline
         \textbf{standard MPC} & $10$ & $0.2$ & & & $5.9\cdot10^3$  \\
         standard MPC & $13$ & $0.2$ & & & $5.9\cdot10^3$  \\
         standard MPC & $16$ & $0.2$ & & & $5.9\cdot10^3$ \\
         standard MPC & $8$ & $0.4$ & & & $6.0\cdot10^3$  \\
         NUSH MPC \cite{TanTippettBao2015} & $10$ & $0.2$ & 3 (det.) & 0.4 & $5.9\cdot10^3$  \\
         \textbf{gran. MPC \cite{BaethgeLuciaFindeisen2016}} & $10$ & $0.2$ & $6$ (cor.) & $0.2$ & $5.6\cdot10^3$   \\
         \textbf{proposed MPC} & $10$ & $0.2$ & $3$ (cor.) & $0.4$ & $5.6\cdot10^3$
    \end{tabular}
    \caption{Comparison of MPC setups: short-term horizon (detailed prediction model), long-term horizon (detailed or coarse prediction model), sampling time, and cost.}
    \label{tab:params}
\end{table}


As shown, the decision variables vary between the three methods. While the horizon of the proposed method and MPC with models of different granularity covers the same horizon, less decision variables are necessary for the proposed approach.

\subsection{Simulation Results}

In this section we will compare the simulation results of the three methods. Each simulation was run for 50 iterations. We will first focus on the individual simulations and then investigate the overall result.

\begin{figure}
    \centering
    \includegraphics[width = \columnwidth]{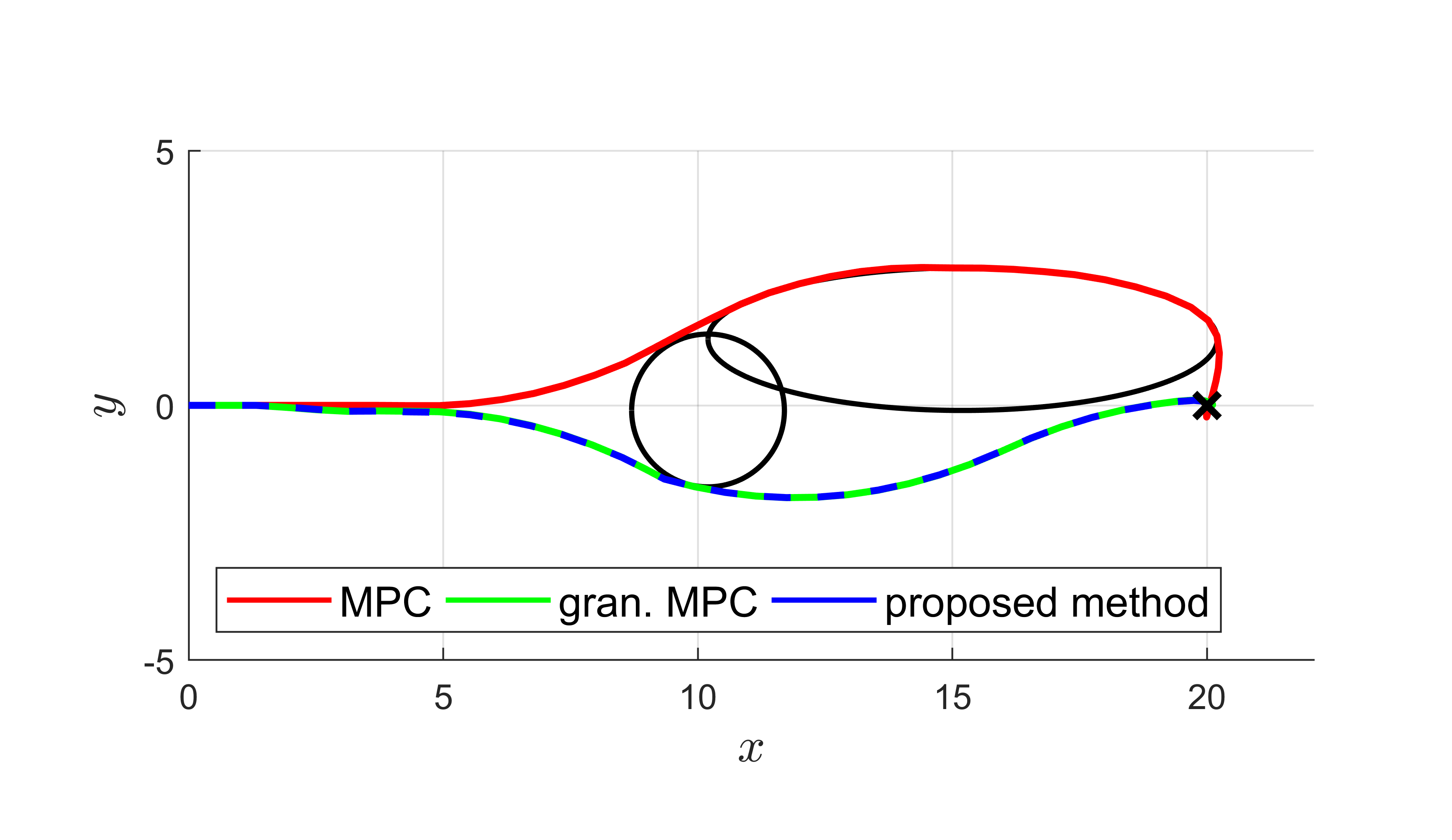}
    \caption{Simulation Results: the standard MPC controller (red) chooses a longer path due to the shorter horizon, while both other approaches find the shorter path.}
    \label{fig:simus}
\end{figure}

The simulation results of the individual controllers are illustrated in Figure~\ref{fig:simus}. As the center of the circular obstacle is set slightly below $y = 0$, the standard MPC controller moves the robot towards the top. If only the circular obstacle were present, this would be the behavior with the lowest cost. However, due to the short horizon the ellipsoidal obstacle is only detected later. As the cost would be larger to change the path, the robot continues the longer path. Both the MPC with models of different granularity and the proposed MPC scheme detect the ellipsoidal obstacle before deciding on a path. Therefore, both methods select the shorter path below the circular obstacle, resulting in lower overall costs.

The overall cost $V^*$ for each simulation run is analyzed by comparing the real cost which occurred for each step, i.e.,
\begin{IEEEeqnarray}{c}
V^* = \sum_{k=0}^{49} l_\text{s}(\bm{x}_{k+1},\bm{u}_k)
\end{IEEEeqnarray}
according to \eqref{eq:cost_s}. The overall results are shown in Table~\ref{tab:params}. The standard MPC with $N = 10$ has the lowest computational effort, the average computation time per iteration is $\SI{0.27}{\second}$ ($100\%$). However, as described before, the shorter horizon results in higher costs, as the longer path is chosen, illustrated by the increased cost compared to the other two methods. In this example, the computational effort of the proposed method is $\SI{0.41}{\second}$ ($151\%$). Eventually, we compare the proposed method with MPC with models of different granularity. While the costs are equal, the proposed method reduces the computational effort by 33\% compared to MPC with models of different granularity ($226\%$).

All three controllers reach the target state eventually, however, cost and computational effort vary. While the proposed method proved to be beneficial here, this is highly scenario dependent. It will be of interest to apply the proposed MPC scheme to more challenging automated vehicle scenarios, considering dynamic obstacles with uncertain behavior \cite{CesariEtalBorrelli2017, BruedigamEtalWollherr2018b}.

\section{Conclusion}
\label{sec:conclusion}

In this paper, we proposed an MPC scheme that combines a detailed model with smaller sampling time and an approximated, coarse model with larger sampling time. The presented method allows to plan precisely on a short-term horizon while still considering long-term goals by improving the cost-to-go. The coarse model combined with increased sampling time allows reduced computational effort.

While recursive feasibility is guaranteed, stability is still an issue, which could be addressed using dissipativity theory, similar to showing stability for MPC with a non-uniformly spaced horizon.

\section*{Acknowledgement}

The authors thank Philipp Bohlig for discussions on MPC with move blocking and a non-uniformly spaced horizon.

\bibliography{./references/Dissertation_bib}
\bibliographystyle{unsrt}

\end{document}